\newcommand\aj{{AJ\,}}%
\newcommand\apj{{ApJ\,}}%
\newcommand\apjl{{ApJ\,}}%
\newcommand\aap{{A\&A\,}}%
\newcommand\nat{{Nature\,}}%

\documentclass[graybox, natbib, footinfo]{svmult}

\usepackage{mathptmx}       
\usepackage{helvet}         
\usepackage{courier}        
\usepackage{type1cm}        
\usepackage{makeidx}         
\usepackage{graphicx}        
\usepackage{multicol}        
\usepackage[bottom]{footmisc}

\makeindex             

\begin{document}

\title*{Open clusters: probes of galaxy evolution and bench tests of stellar models}
\author{Maurizio Salaris}
\institute{Maurizio Salaris \at Astrophysics Reaserch Institute, Liverpool John Moores University, 
IC2, Liverpool Science Park, 
146 Brownlow Hill, 
Liverpool L3 5RF,
United Kingdom\email{M.Salaris@ljmu.ac.uk}}
%
%
\maketitle

\abstract{Open clusters are the only example of single-age, single initial chemical composition populations in the Galaxy, and they 
play an important role in the study of the formation and evolution of the Galactic disk. In addition, 
they have been traditionally employed 
to test theoretical stellar evolution models. A brief review of constraints/tests of white dwarf models/progenitors, and 
rotating star models based on Galactic open clusters' observations is presented, introducing also recent contributions of 
asteroseismic analyses.}

\section{Introduction}
\label{sec:1}

Star clusters had traditionally played --and continue to play-- a fundamental role as tools to both investigate the mechanisms 
of galaxy formation and evolution, and test theoretical stellar evolution models.
For example, the timescale for the formation of the different Galactic populations 
can be investigated by means of stellar age dating. The most reliable stellar ages are obtained for 
the globular clusters in the halo, thick disk and bulge, 
and the open clusters (OCs) in the thin disk. 
In case of OCs, it is possible to employ 
techniques like gyrochronology or the lithium depletion boundary on young OCs,  
in addition to classical methods based on isochrone and colour-magnitude-diagrams (CMD).

Given the recent 'downgrading' of individual globular clusters from SSPs to ensembles of almost coeval stellar 
populations with a largely varying abundances of specific elements (C, N, O, Na, Mg, Al and He), OCs 
remain the only example of pure SSP in the Galaxy. 
There are about 1500 known Galactic OCs, distributed at galactocentric distances (${\rm R_{GC}}$) 
between $\sim$5 and 20 kpc, and ages between a few Myr up to $\sim$10~Gyr.
The old-age tail of this distribution is particularly important for Galaxy formation studies.  
In general, OCs are expected to be disrupted 
easily by encounters with massive clouds in the disk; however, the most massive OCs or those with orbits 
that keep them far away from the Galactic plane for most of their lifetimes are expected to survive for longer periods 
of time. These old objects are therefore test particles - in analogy to the GCs - probing the earliest stages of the 
formation of the Galactic disk.


Besides the use of OCs to study the Galaxy, their CMDs provide a snapshot of magnitudes and colours of 
coeval and uniform initial composition stars of different masses at different evolutionary stages. 
CMD analyses plus star counts and spectroscopic studies 
along the various evolutionary sequences provide strong tests/constraints on stellar physics and evolutionary predictions.

The next sections will review briefly some OC-based constraints/tests of white dwarf models/progenitors, and 
rotating stellar models, introducing also the recent constributions of asteroseismic observations.

\section{White Dwarfs}
\label{sec:2}

White dwarfs (WDs) are the last evolutionary phase of stars with initial mass smaller than about 10-11${\rm M_{\odot}}$. 
The large majority of WDs, i.e. those with progenitor mass below 6-7${\rm M_{\odot}}$ are made of an electron degenerate core of 
carbon and oxygen. More massive WDs have a oxygen and neon cores.
Given that most stars are or will become WDs, plus the existence of a 
relationship between their cooling time and luminosity, and their long timescales, 
WDs are attractive candidates to unveil the star formation history of the Galaxy. 
Due to their proximity compared to globular clusters, and their large range of ages, 
Galactic OCs are perfect systems to employ WDs as cosmochronometers, and study their properties.

\subsection{Initial-final mass relation}

The initial-final mass relation (IFMR) for low- and intermediate-mass stars is an important input for many astrophysical 
problems. Given the initial main sequence mass of a formed star, the IFMR provides the expected mass during its final 
WD cooling stage, and is an estimate of the total mass lost by the star during its evolutionary history. A 
correct assessment of the IFMR is very important when predicting, for example, the chemical evolution history of stellar 
populations, or their mass-to-light ratio (defined as the ratio of the mass of evolving stars plus remnants--WDs, neutron 
stars and black holes--to the integrated luminosity of the population), and in general for any problem related to the origin 
and evolution of gas in stellar populations. 

The IFMR depends on the competition between surface mass loss, growth of the CO core due to shell He-burning during the asymptotic giant branch phase, 
and the mixing
episodes between envelope and intershell region, that limit the outward
(in mass) movement of the He-burning shell.
Due to our imperfect knowledge of 
mass loss processes during the asymptotic giant branch and post asymptotic giant branch phases, and details of the mixing during the thermal pulses,  
we cannot predict accurately the mass of a WD
produced by a progenitor with a given initial mass. Observational constraints are therefore absolutely necessary. 

OCs have traditionally provided the observational data to determine semiempirically the IFMR, starting from \cite{weide}.
These determinations based on cluster WDs work as follows: after detection, spectroscopic estimates of the WD surface gravity g 
and ${\rm T_{eff}}$ are needed. For a fixed g- ${\rm T_{eff}}$ pair, interpolation within a grid of theoretical WD models covering a range of masses 
provides the final WD mass and the cooling age of the WD. Independent theoretical isochrone fits to the turn-off luminosity in 
the cluster CMD provide an estimate of the cluster age. The difference between cluster and WD cooling ages is equal 
to the lifetime of the WD progenitor from the main sequence until the start of the WD cooling (that is essentially the same as the lifetime 
at the end of central He-burning). Making use of mass-lifetime 
relationships from theoretical stellar evolution models (wihout including the short-lived 
asymptotic giant branch phase), the initial progenitor mass is derived.

%
\begin{figure}[b]
\includegraphics[scale=.5000]{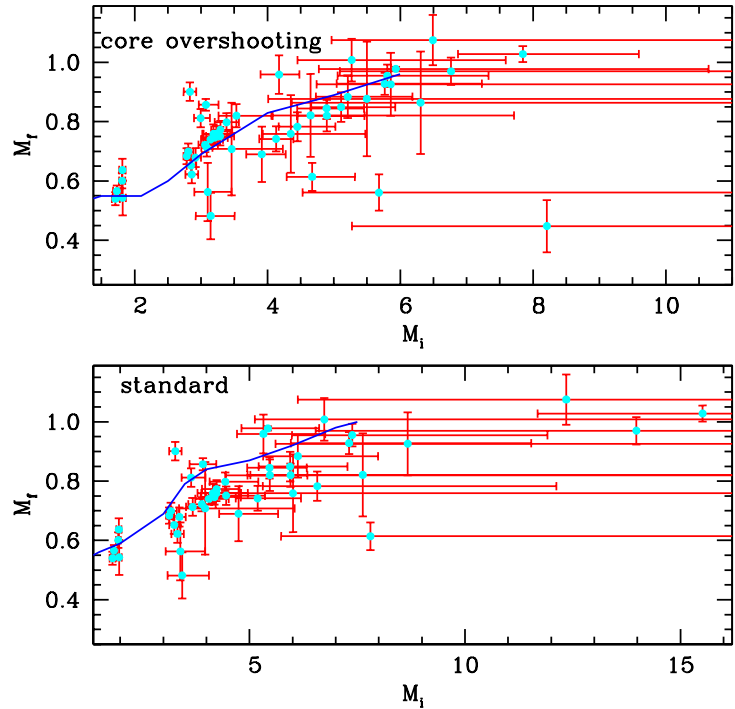}
\caption{Semiempirical IFMR determined with models with and without main sequence convective core overshooting, as labelled.
The solid lines display theoretical IFMRs obtained from synthetic AGB models (see text for details)}
\label{fig:ifmr}       
\end{figure}

An important issue here is the knowledge of the initial chemical 
composition of the cluster, given the strong dependence of the derived ages on the assumed metallicity 
of the models.
The recent detailed analysis by \cite{sal:09} based on 10 OCs, 
have shown that the uncertainty in the final WD mass is dominated by observational errors, whilst 
the uncertainty in the initial mass has multiple reasons. On the one hand cluster chemical composition  
and isochrone details influence the cluster age and thus the progenitor mass; but also 
the uncertainty in the WD cooling age can sometimes be the dominant factor. 
None of the WDs employed in current IFMR determinations is close to the Chandrasekhar mass, 
not even the progeny of the more massive intermediate mass stars.
Figure~\ref{fig:ifmr} displays the empirical IFMR determined by \cite{sal:09} employing  BaSTI 
stellar isochrones with and without main sequence core overshooting from \cite{basti}. Overimposed are predicted 
relationships from BaSTI synthetic AGB models by \cite{bastiAGB}, with and without main sequence convective core overshooting. Clearly, 
stellar models without convective overshooting during core hydrogen burning lead to internal inconsistencies in the semiempirical IFMR.

\subsection{WD ages}

There are two main age indicators for stellar populations: the main sequence turn-off and the termination of the WD cooling sequence.
Open clusters provide the ideal environment for the test/calibration of these two clocks. 
Although generally WD and turn-off ages are consistent within the error bars for the  OCs where 
both indicators have been applied \citep{vonHippel2005}, observations of the WD cooling sequence of the old open cluster NGC6791 
has revealed a few surprises. 

%
\begin{figure}[b]
\includegraphics[scale=.4500]{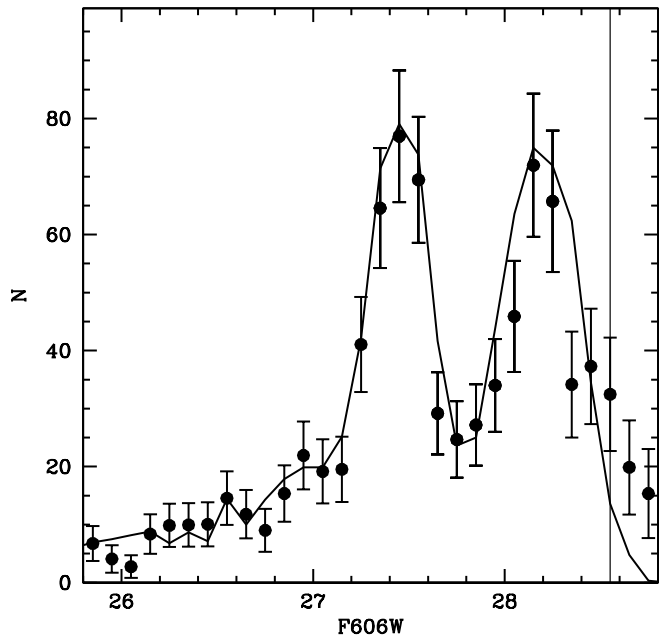}
\caption{WD luminosity function of NGC6791 (filled circles), and the best-match theoretical counterpart (solid line) 
with a $\sim$30\% fraction of WD+WD binaries (see text for details)}
\label{fig:n6791}       
\end{figure}

A deep CMD by \cite{bedin} has provided a well populated WD luminosity function (LF) that reaches the end of the cluster cooling sequence. 
The LF, displayed in Fig.~\ref{fig:n6791}, displays 
a peak and sharp cut-off at low luminosities, caused by the finite age of the cluster (hence the finite WD cooling time), 
plus an unexpected  secondary peak at higher luminosities, never observed in any other OC cooling sequence so far.

The cluster age derived from the magnitude of the cut-off of the LF appeared to be about $\sim$2~Gyr younger than the turn-off age, when calculated 
with standard WD cooling models. 
As shown conclusively by \cite{diff} the inclusion of ${\rm ^{22}Ne}$ diffusion in the core, in the liquid phase --a physical process never 
included before in standard WD model calculations-- slows down the cooling and 
can explain the discrepancy with the turn-off ages. The ${\rm ^{22}Ne}$ mass fraction in the core of NGC6791 WDs should be about 4\% by mass, 
essentially equal to the total mass fraction Z of this metal rich OC ([Fe/H]$\sim$0.3-0.4). 
At the solar metallicities typical of the Galactic OCs, the amount of  ${\rm ^{22}Ne}$ is not large enough to cause appreciable delays 
in the WD cooling, hence the lack of discrepancy between turn-off and WD ages for other OCs. 
This result highlights very clearly the power of using OCs as tools to test and improve stellar evolution calculations.

The secondary peak at higher luminosities is more puzzling, and it could be produced by a population of unresolved binary white dwarfs.
As shown by \cite{bedinbin} $\sim$30\% of unresolved WD+WD systems arising from a $\sim$50\% initial fraction of binary systems 
can reproduce both height and magnitude of the secondary peak (see Fig.~\ref{fig:n6791}).
An alternative explanation put forward by \citet{Hansen2005} is the presence of massive He-core WDs (mass essentially equal to the core mass 
at the He-flash along the RGB). This idea is supported by the fact that NGC 6791 contains a non-negligible number of blue He-burning stars 
that have very little mass in their envelopes, having lost nearly all of the envelope during their red giant branch evolution. 
This explanation requires a certain fine-tuning of the initial-final mass relation for these He-core objects, and an overall very large amount 
of mass lost along the red giant branch.

\section{Rotation}

In the regime of solar-like stars, it is long known that 
stellar rotation periods increase approximately as the square root of age, due mainly to 
mass and angular momentum loss in a magnetized wind, see e.g., \cite{skumanich}. 
Theoretical predictions of the rotational evolution of stars are however difficult, because 
the theory of angular momentum evolution in stars is very complex, for one has to understand not only the 
initial distribution of angular momentum, but also its transport in stellar interiors and wind losses. 

Observations of the evolution of rotational periods in nearby OCs provide 
important clues about how to use the evolution of 
rotational properties as a clock for low mass main sequence field stars, whose ages would be extremely difficult if not impossible to measure from 
their position in CMDs. These empirical results, in turn, can be used to calibrate the rotational evolution of stellar models.

Figure~\ref{fig:rot} shows measurements of Period ($P$) against colour $(B-V)$ of samples of main sequence stars (in the range 
between $\sim$0.6 and $\sim$1.4${\rm M_{\odot}}$) in the 
$\sim$600~Myr old Hyades, and in the $\sim$120~Myr old M35 OCs.
It is easy to appreciate the large period spread of $P$ values 
at fixed colour in M35, with two well defined sequences. One sequence of fast rotators 
with period $P<$1~day, independent of colour --denoted as sequence $C$ in \cite{barnes}-- or  
convective sequence, in the assumption that these objects lack large scale dynamos and are inefficient at slowing down 
their rotation; 
a diagonal sequence of faster rotating/warmer stars and slower rotating/cooler stars (sequence $I$, or interface, given 
the theoretical expectation that these stars are producing their magnetic flux near the convective-radiative interface). 
A comparison of the two clusters suggests that by the age of the Hyades almost all stars along sequence $C$ have 
moved onto sequence $I$. The stars populating the gap between these two sequences in the younger cluster are interpreted 
as objects in transition from the $C$ to the $I$ sequence. 
The colour-P diagrams also suggest that the dependence of $P$ on colour along the $I$ sequence is the same in both clusters, hence 
the value of the rotation period $P$ along this sequence can be expressed as:

%
\begin{figure}[b]
\includegraphics[scale=.5]{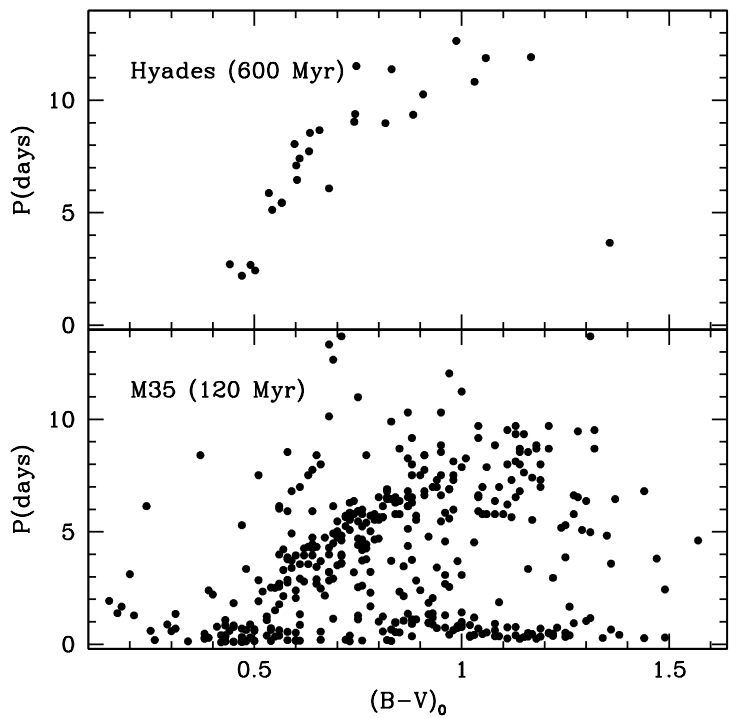}
\caption{Colour-$P$ diagram for a sample of main sequence stars in the Hyades and M35 OCs (see text for details)}
\label{fig:rot}       
\end{figure}

\begin{equation}
P={\rm f}(B-V) \ {\rm g(t)} 
\end{equation}

with ${\rm f}(B-V)=a \ [(B-V)_0 - c]^b $ and ${\rm g(t)}={\rm t}^n$, as proposed by \cite{barnes}.
The recent determinations of the coefficients $a, b, c, n$ by \cite{meibom} provide 
$a=0.770\pm0.014$, $b=0.553\pm0.052$, $c=0.472\pm0.027$, based on the $I$ sequence in M35. The exponent 
$n$ is determined by ensuring that the colour dependence gives the solar rotation period at the solar age, that  
gives $n=0.519\pm 0.007$.
Age determinations based on this so-called 'gyrochronology' rely on fitting the $I$ sequence rotational isochrone determined by 
Eq.~\ref{fig:rot} with age as a free parameter, to individual field stars --or clusters, where one can 
determine the position of the $I$ sequence even in the presence of fast rotators-- in the colour-period diagram.
This calibration is based on OCs younger than 1~Gyr, with a time 
dependence covering ages up to $\sim$5~Gyr, based just on the Sun. 
Very recently \cite{meibom11} have used photometry from the {\sl Kepler} mission, and ground based spectroscopy to determine 
the colour-$P$ diagram of the $\sim$1~Gyr old OC
NGC 6811, that shows clearly the $I$ sequence.
This improves the calibration of gyrochronology by extending the age baseline 
of the reference clusters. 

These results are not only important for gyrochronology, but 
can be used also to constrain the rotational evolution of cool stars during the main sequence phase. 
Let's remember that rotation plays a crucial role in stellar structure and its evolution, for it influences the evolutionary tracks in the CMD 
through transport processes which induce rotational mixing of chemical species and the redistribution of angular momentum. 
In turn, stellar evolution affects the rotational properties. 

Additional information on the rotational properties of the deep interior would also 
help to better understand the effect of rotation on stellar evolution, and recently 
rotational splittings of dipole mixed modes were measured by \cite{mosser} in about 300 red giants observed during more than two years with {\sl Kepler}. 
The measured splittings provide an estimate of the mean core rotation period, and reveal that along the  
the red giant branch the period increases more slowly than expected in case 
of homologous spinning down at constant total angular momentum. 
Angular momentum is transferred from the core to the envelope, but a strong differential rotation profile takes place 
during the red giant branch ascent. 
Rotation periods are larger for red clump stars compared to red giant branch objects, implying a transfer of angular momentum 
from the rapidly rotating core to the slowly rotating envelope, however 
the mechanism responsible for the redistribution of angular momentum spins down the mean core rotation 
with a time scale too long for reaching a solid body rotation. 

\begin{acknowledgement}
I wish to thank the organizers for their kind invitation and the organization of the workshop in such an enchanting place.
\end{acknowledgement}
%

%


\end{document}